\documentclass[fleqn,usenatbib]{mnras}

\usepackage[T1]{fontenc}

\DeclareRobustCommand{\VAN}[3]{#2}
\let\VANthebibliography\thebibliography
\def\thebibliography{\DeclareRobustCommand{\VAN}[3]{##3}\VANthebibliography}

\usepackage{graphicx}	
\usepackage{amsmath}	
\usepackage{amssymb}	
\usepackage{siunitx,xcolor,tikz}

\newcommand{\ud}{\mathrm{d}}   
\newcommand{\ue}{\mathrm{e}} 
\definecolor{lime}{HTML}{A6CE39}
\DeclareRobustCommand{\orcidicon}{\hspace{-3mm}
	\begin{tikzpicture}
	\draw[lime, fill=lime] (0,0) 
	circle [radius=0.16] 
	node[white] {\hspace{0.1mm}{\fontfamily{qag}\selectfont \tiny ID}};
	\draw[white, fill=white] (-0.07,0.1) 
	circle [radius=0.01];
	\end{tikzpicture}
	\hspace{-5mm}
}

\foreach \x in {A, ..., Z}{\expandafter\xdef\csname orcid\x\endcsname{\noexpand\href{https://orcid.org/\csname orcidauthor\x\endcsname}
		{\noexpand\orcidicon}}
}

\title[Implications of EDGES]{Implications of the cosmological 21-cm absorption profile for high-redshift star formation and deep JWST surveys}
\author[Mittal and Kulkarni]{
Shikhar Mittal\ \ \orcidA{}\ \ \thanks{E-mail: shikhar.mittal@tifr.res.in}
and Girish Kulkarni\ \ \orcidB{}\ \ \thanks{E-mail: kulkarni@theory.tifr.res.in}
\\
Tata Institute of Fundamental Research, Homi Bhabha Road, Mumbai 400005, India
}

\DeclareSIUnit\parsec{pc}
\defcitealias{Mittal_2021}{M22}

\newcommand{\mx}{\mathcal{M}_\mathrm{X}}
\newcommand{\ma}{\mathcal{M}_\alpha}
\newcommand{\mstar}{\mathcal{M}_\star}

\newcommand{\lya}{Ly~$\alpha$}

\date{Accepted 2022 July 07. Received 2022 July 07; in original form 2022 March 17}

\pubyear{2022}
\usepackage{newtxtext,newtxmath}
\begin{document}
\label{firstpage}
\pagerange{\pageref{firstpage}--\pageref{lastpage}}
\maketitle

\begin{abstract}
Apart from its anomalously large depth, the cosmological 21-cm absorption signal measured by the EDGES collaboration also has a shape that is distinctly different from theoretical predictions. Models with non-traditional components such as super-adiabatic baryonic cooling or an excess radio background explain the depth of the observed profile, but still conspicuously fail to explain its shape. In this paper, we quantify the requirements imposed by the EDGES measurement on sources of \lya\ and X-ray photons in the presence of excess radio background at cosmic dawn. In extreme cases, the \lya\ and X-ray emissivities require to be enhanced by up to an order of magnitude relative to traditional models. Furthermore, this enhancement needs to be active only for a short duration. We find that under conventional assumptions for the cosmic star formation rate density, standard stellar populations are incapable of meeting these conditions. Only highly unusual models of massive metal-free stars seem to provide a possible mechanism. Conversely, if the sources of \lya\ and X-ray photons are compelled to have standard properties, the EDGES measurement puts strong demands on the cosmic star formation rate density. This provides interesting falsifiable predictions for high-redshift galaxy surveys enabled by \textit{James Webb Space Telescope} (\textit{JWST}). We derive predictions for galaxy UV luminosity functions and number densities, and show that a deep \textit{JWST} survey with a limiting UV magnitude of $m_\mathrm{UV,lim}=32$ would potentially be able to rule out the predictions enforced by the EDGES measurement.
\end{abstract}

\section{Introduction}

The absorption signal in the low-frequency spectrum of the sky reported by the EDGES collaboration \citep{Bowman} has been widely considered to potentially have a cosmological origin \citep{B18, Mirocha19}. But it is also famously known that the amplitude of this signal, $\Delta T_{\mathrm{b}}=0.5^{+0.5}_{-0.2}\,$K at $z\approx17.2$, is larger than the expected cosmological value by a factor of two. This anomalously large depth has received considerable attention in the literature. Many studies explain the depth either by postulating excess cooling of cosmic hydrogen due to novel physical mechanisms \citep[eg.][]{Meiksin_2020, Li, mathur2021}, or by bringing into play a part of the excess radio background (ERB) observed at relevant frequencies \citep[eg.][]{Mittal_2022}. Some studies invoke even more exotic scenarios, such as a modified Hubble parameter at cosmic dawn \citep{Hill_2018, costa, Wang2018, xiao}, new coupling terms in the spin temperature \citep{Widmark_2019, Lamb}, or special classes of dark energy and dark matter models \citep{Upala}.


Still, apart from the large depth, which in this work we explain by incorporating excess radio background, the absorption profile measured by EDGES also has peculiarly sharp bounds, at $z\sim 20$ and $z\sim 15$, and a surprisingly flat bottom \citep{Bowman}. At the minimum, these unusual features in the shape of the signal seem to require that a cosmic Ly~$\alpha$ background must get established within 25\,Myr from $z\sim 21$ to 19 and that the cosmic gas content be heated above the background temperature within 50\,Myr from $z\sim 16$ to 14. This aspect of the EDGES measurement has received relatively less attention in the literature. For example, \citet{Kaurov_2018} argue that the rapid Ly~$\alpha$ production required by the falling edge of the signal suggests star formation was dominated by rapidly growing rare and massive haloes. \citet{Venno} claimed that such a falling feature could be explained by warm dark matter models and put upper and lower bounds on the mass of their particles. Population~III motivated models by \citet{mebane} naturally achieved a reasonably flat trough, though the rise out of absorption was not too rapid. In this work we aim to interpret the overall shape of the EDGES signal, capturing not just its timing, width or depth but also the sharp drop and rise.

It is possible that the signal measured by EDGES is not of a cosmological origin. Indeed, \citet{Hills} argued against a cosmological origin of this signal by pointing out that the EDGES inference results in unphysical properties for the ionosphere and/or unexplained structure in the foreground spectrum. The inferred absorption profile is also sensitive to the order of the polynomial model used for the foregrounds, without a means to choose the optimum order. More recently, \cite{saras3} used observations made with the SARAS~3 (Shaped Antenna measurement of the background RAdio Spectrum) experiment to rule out an astrophysical origin of the EDGES measurement at 93.5 per cent confidence. Still, in spite of these new developments, an investigation into the shape of the absorption profile measured by EDGES continues to be of interest. For instance, the new findings by \cite{saras3} seem to remain in need of corroboration. First, the frequency band of SARAS~3 covers only part of the the EDGES absorption profile, with the rising part of the signal noticeably excluded. This makes it likely that any signal consistent with the EDGES measurement gets partly or wholly fitted out by the sixth-order polynomial foreground model used in the SARAS~3 analysis. [Although it is important to note that \cite{saras3} have argued against this possibility by showing that even with a partial coverage of 55--85\,MHz, the EDGES signal should be detected by SARAS~3 against a sixth-order polynomial at more than two-sigma confidence.] Second, the choice of a sixth-order polynomial foreground model by SARAS~3 is as problematic as the use of the fifth-order linearised foreground model in the EDGES analysis by \citet{Bowman}.  As \citet{Hills} pointed out, it is surprising that such higher-order models are necessary to describe the foreground for which third-order polynomials are usually sufficient \citep{Bernardi_2015}.  Third, it is remarkable that with a fifth-order polynomial foreground, both SARAS~3 and EDGES report residuals that are roughly identical in the rms amplitude. Until a reliable explanation is found for this residual, the possibility of its astrophysical origin remains open. It is also interesting to note that \citet{Bowman} found the shape of the inferred absorption signal in the EDGES data to be robust against different hardware modifications, data processing pipelines, and calibration techniques \citep{Mahesh_2021}. Newer high-band measurements by the EDGES collaboration also seem to confirm this \citep{Monsalve_2019}.  Further, it is important to note that more experiments are currently being set up to verify the EDGES and SARAS~3 observation.  These include Large Aperture Experiment to Detect the Dark Ages \citep[LEDA,][]{Bernardi_2015, Bernardi_16, Price}, Probing Radio Intensity at high-Z from Marion \citep[PRIzM,][]{philip}, Radio Experiment for the Analysis of Cosmic Hydrogen \citep[REACH,][]{Eloy}, Sonda Cosmol{\'o}gica de las Islas para la Detecci{\'o}n de Hidr{\'o}geno Neutro \citep[SCI-HI,][]{Scihi} and the Cosmic Twilight Polarimeter \citep[CTP,][]{Nhan_2017, Nhan_2019}. If only conventional seven-parameter models are used for the cosmological signal in the analysis of these experiments, the analysis of data from these upcoming experiments runs the risk of missing cosmological information. Finally, the successful launch of \textit{James Webb Space Telescope} (\textit{JWST}) has opened a window of opportunity for a rest-frame UV detection of sources of radiation at cosmic dawn \citep{cowley, Williams_2018, kauff, Mirocha19}. A quantification of the implications of 21-cm experiments for \textit{JWST} surveys is therefore timely. 

The major standard astrophysics components affecting the global 21-cm signal for the period of cosmic dawn are the cosmic star formation rate (SFR), the Ly~$\alpha$ coupling via the Wouthuysen--Field effect \citep{Wouth, Field}, and gas heating due to X-ray emission. This astrophysics is conventionally described by 4 to 7 parameters \citep{Mirocha_2015, Cohen17, Monsalve_2019, mebane, Atrideb}, usually chosen to describe the star formation efficiency (SFE), minimum mass of dark matter haloes that host star formation, the normalization and spectral index for the Ly~$\alpha$ spectral energy distribution (SED), and the normalization and spectral index for the X-ray SED. In this work we explore extensions to these models if they are to also capture the shape of the absorption profile.

We start by briefly mentioning key equations describing the 21-cm signal in Section~\ref{sec:basics}. In Section~\ref{sec:new_models}, we discuss the astrophysical requirements demanded by the shape of the observed 21-cm absorption feature. The plausibility of these requirements is discussed in Section~\ref{sec:plaus}, and Section~\ref{sec:implic} presents the consequent implications for high-redshift galaxies. We discuss and summarise our findings in Section~\ref{sec:conc}. Our cosmological parameter values are $\Omega_{\mathrm{m}}= 0.315$, $\Omega_{\mathrm{b}}=0.049$, $\Omega_\Lambda = 0.685$, $h=0.674$, $Y_{\mathrm{p}}=0.25$, $T_0=\SI{2.725}{\kelvin}$, $\sigma_8 = 0.811$ and $n_{\mathrm{s}} = 0.965$ \citep{Fixsen_2009, Planck}, where $T_0$ and $Y_{\mathrm{p}}$ are the CMB temperature measured today and primordial helium fraction by mass, respectively.

\begin{figure*}
\centering
\includegraphics[width=1\textwidth]{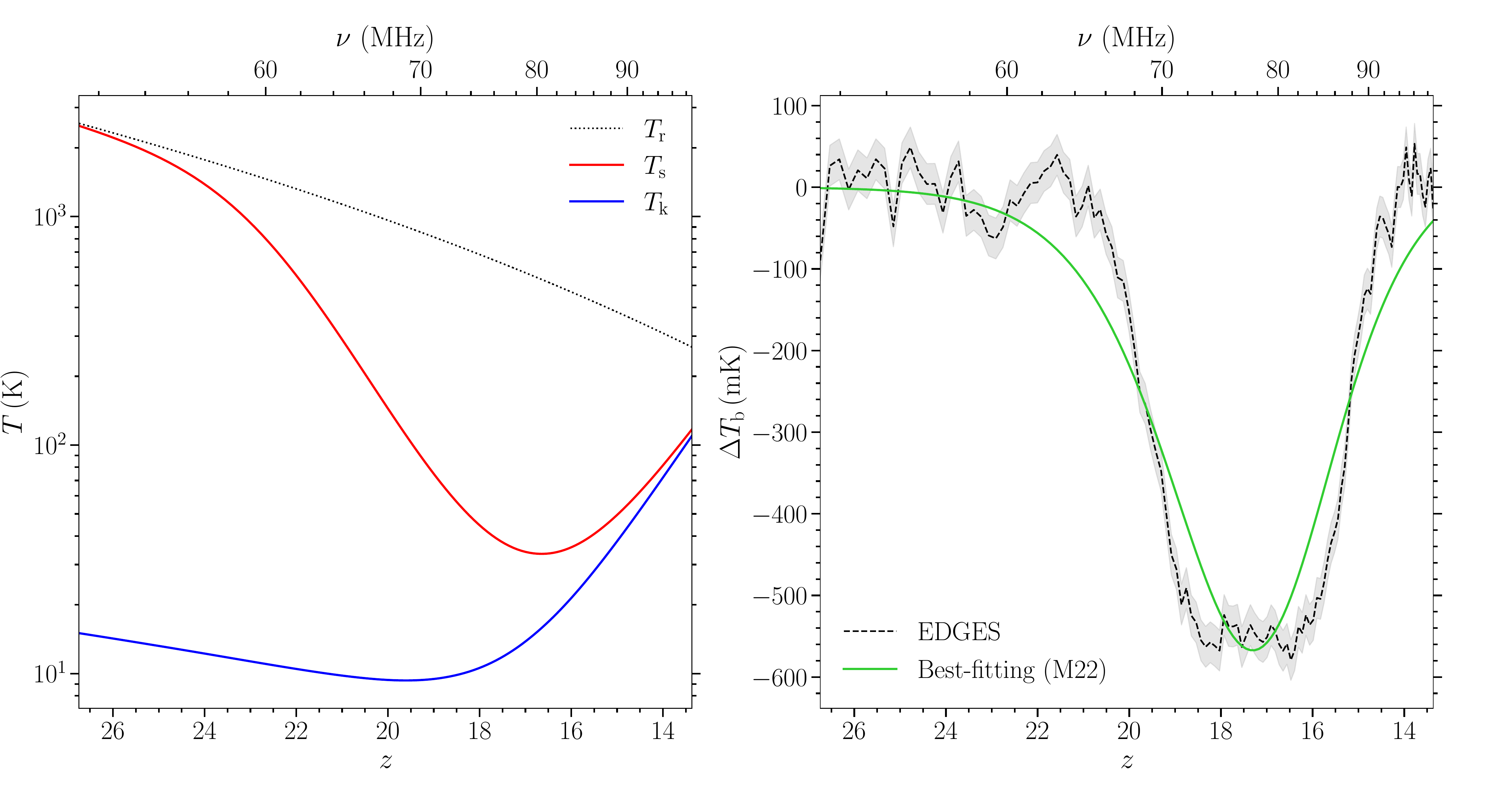}
\caption{Our fiducial seven-parameter model for the 21-cm signal from \citetalias{Mittal_2021}.  The left panel shows the evolution of the radio background temperature (dotted black curve), the hydrogen spin temperature (solid red) and the gas temperature (solid blue). The right panel shows the best-fitting model (solid green) in comparison with the EDGES measurement (dashed black), with a $\SI{50}{\milli\kelvin}$ uncertainty. The best-fitting values of the model parameters are given in Table~\ref{Tab1}. The goodness-of-fit of this model is $\chi^2/\mathrm{dof}=2.20$ (right tail $p$-value $8.7\times 10^{-13}$). The rms value of the residuals is $\SI{69}{\milli\kelvin}$.}\label{Fig1}
\end{figure*}

\begin{table*}
\centering
\caption{Best-fitting parameter values for the seven-parameter model shown in Figure~\ref{Fig1}.}\label{Tab1}
\begin{tabular}{lll}
\hline
Parameter  & Description & Value\\ \hline
$f_{\star}$ & Star formation efficiency & 0.1\\
$T_{\mathrm{vir,min}}$ & Minimum virial temperature of star-forming haloes & $\SI{e4.25}{\kelvin}$\\
$f_\alpha$ & Normalization of the Ly~$\alpha$ background & $0.985$\\
$E_0$ & Lower X-ray energy cut-off & $\SI{200}{\electronvolt}$\\
$w$ & X-ray spectral index & 1.5\\
$f_\mathrm{X}$ & Normalization of the X-ray background & $2.379$\\
$\zeta_{\mathrm{ERB}}$ & Strength of the ERB & 0.030\\ \hline
\end{tabular}
\end{table*}

\section{The 21-cm absorption profile}
\label{sec:basics}

The 21-cm signal is a differential brightness temperature measured at low frequencies against the radio background. This can be written as \citep{Furlanetto, Pritchard_2012, B16, Mesinger19, Liu_2020}
\begin{equation}
  \Delta T_{\mathrm{b}}=\frac{T_\mathrm{s}-T_\mathrm{r}}{1+z}\left(1-\ue^{-\tau_{21}}\right)\,,
  \label{eqn:delta_tb}
\end{equation}
where $\tau_{21}$ is the optical depth for the 21-cm photon \citep{MMR}, $T_\mathrm{s}$ is the hydrogen spin temperature, $T_\mathrm{r}$ is the radio background temperature. For small 21-cm optical depth Equation~\eqref{eqn:delta_tb} becomes
\begin{equation}
\Delta T_\mathrm{b}=27\bar{x}_{\ion{H}{i}}\left(\frac{1-Y_{\mathrm{p}}}{0.76}\right)\left(\frac{\Omega_\mathrm{b}h^2}{0.023}\right)\sqrt{\frac{0.15}{\Omega_\mathrm{m}h^2}\frac{1+z}{10}}\left(1-\frac{T_\mathrm{r}}{T_\mathrm{s}}\right)\si{\milli\kelvin}\,,\label{DeltaT}
\end{equation} 
where $x_{\ion{H}{i}}\equiv n_{\ion{H}{i}}/n_{\text{H}}$ is the ratio of number densities of neutral hydrogen (\ion{H}{i}) and total hydrogen (H). The bar on $x_{\ion{H}{i}}$ represents a global average over the cosmic volume. At the high redshifts that correspond to the cosmic dawn, $H(z)=H_0\sqrt{\Omega_\mathrm{m}(1+z)^3}$, where $H_0$ is the Hubble's constant measured today. The signal is seen in absorption if $\Delta T_\mathrm{b}<0$ and in emission if $\Delta T_\mathrm{b}>0$. The temperature of the radio background is \citep{Feng_2018, Fialkov19}
\begin{equation}
T_{\mathrm{r}}(z)=2.725(1+z)\left[1+0.169\zeta_{\mathrm{ERB}}(1+z)^{2.6}\right]\,,\label{excess}
\end{equation}
where we have assumed a power-law ERB over the CMB with $\zeta_{\mathrm{ERB}}=0.03$. Our ERB model is based on ARCADE2/LWA1 observations \citep{Fixsen_2011, Dowell_2018}. The value of $\zeta_{\mathrm{ERB}}$ quoted is required to explain the depth seen in EDGES observation \citep[hereafter \citetalias{Mittal_2021}]{Mittal_2021}.

\begin{figure*}
\centering
\includegraphics[width=0.8\textwidth]{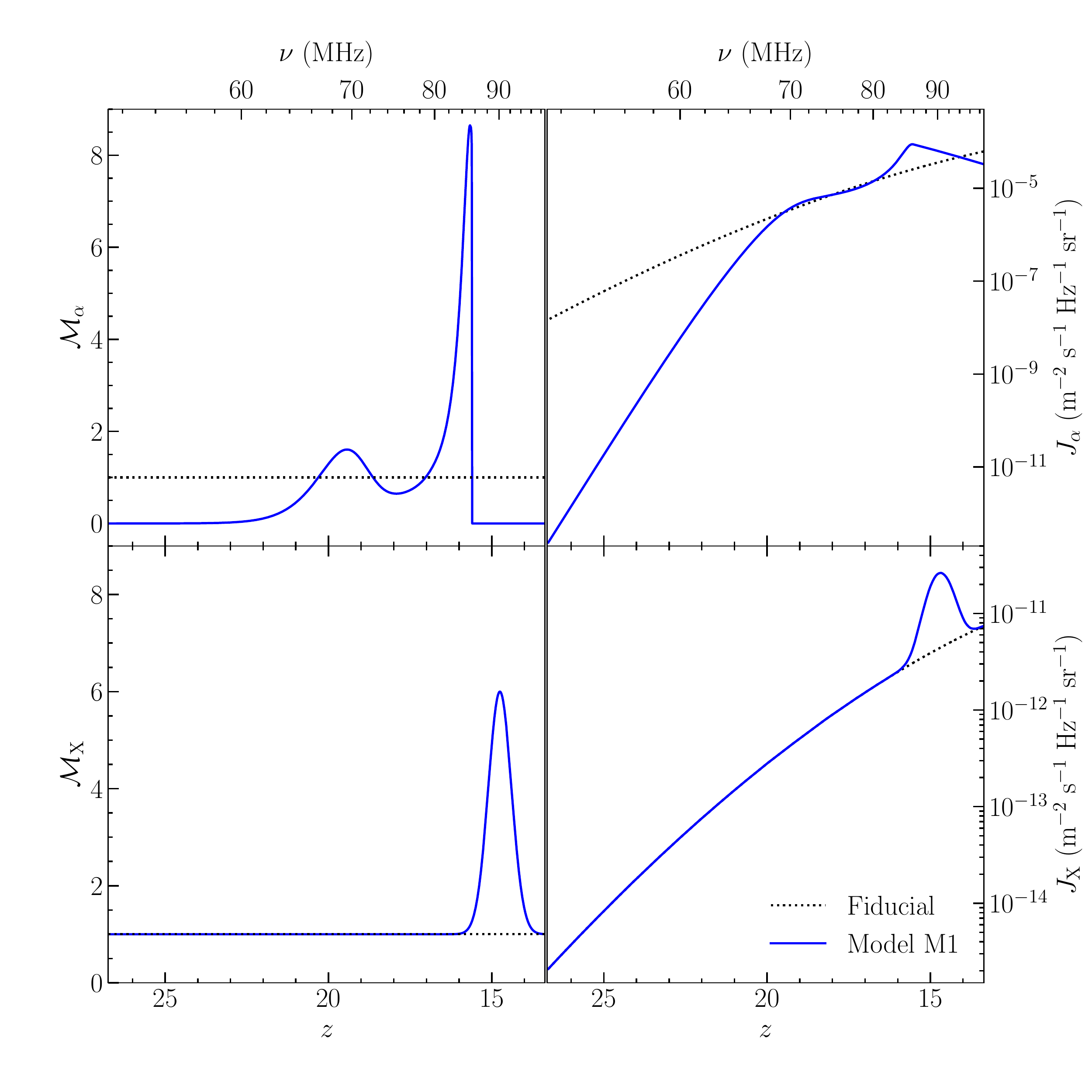}
\caption{Our model~M1. The left-hand side panels show the multiplicative factors $\ma$ and $\mx$ for the \lya\ and X-ray emissivity, respectively, required to fit the flat-bottomed shape of the EDGES measurement. The right-hand side panels show the corresponding background intensities $J_\alpha$ and $J_{\mathrm{X}}$. Dotted curves in all panels denote the fiducial model. This model presents the extreme scenario in which the X-ray emissivity is held as close to the fiducial model as possible. Note that this model explains the flat bottom of the EDGES profile by mean of an evolution in the Ly~$\alpha$ coupling.}\label{Fig2}
\end{figure*}

The spin temperature is given by
\begin{equation}
T_{\mathrm{s}}^{-1}=\frac{T_{\mathrm{r}}^{-1}+ x_\alpha T_\mathrm{k}^{-1}}{1+x_\alpha}\,.\label{Ts}
\end{equation}
where $x_{\alpha}$ is the Ly~$\alpha$ coupling. Note that we have ignored the collisional coupling as it is quite small in the redshift range of our interest \citep{Mittal_2020}.

We have also approximated the colour temperature \citep{Rybicki2006, Meiksin_2006} by the gas temperature. The evolution of the gas temperature $T_{\mathrm{k}}$ is determined by adiabatic cooling and X-ray heating \citep{Mesinger11} so that
\begin{equation}
(1+z)\frac{\ud T_{\mathrm{k}}}{\ud z}=2T_{\mathrm{k}}-\frac{2q_{\mathrm{X}}}{3n_{\mathrm{b}}k_{\mathrm{B}}H(z)}\,,\label{tkeqn}
\end{equation}
where $k_{\mathrm{B}}$ is the Boltzmann constant, $n_{\mathrm{b}}$ is the baryon number density and $q_{\mathrm{X}}$ is the volumetric heating rate by X-rays, which is proportional to the specific intensity $J_{\mathrm{X}}$ of X-ray photons. We ignore other heating/cooling processes such as Compton \citep{Weymann}, Ly~$\alpha$ \citep{Mittal_2020} and change in internal energy due to changing particle number as they are very small compared to X-ray heating at the redshifts of our interest. Further details of our seven-parameter model are given by \citetalias{Mittal_2021}.

\begin{figure*}
\centering
\includegraphics[width=1\textwidth]{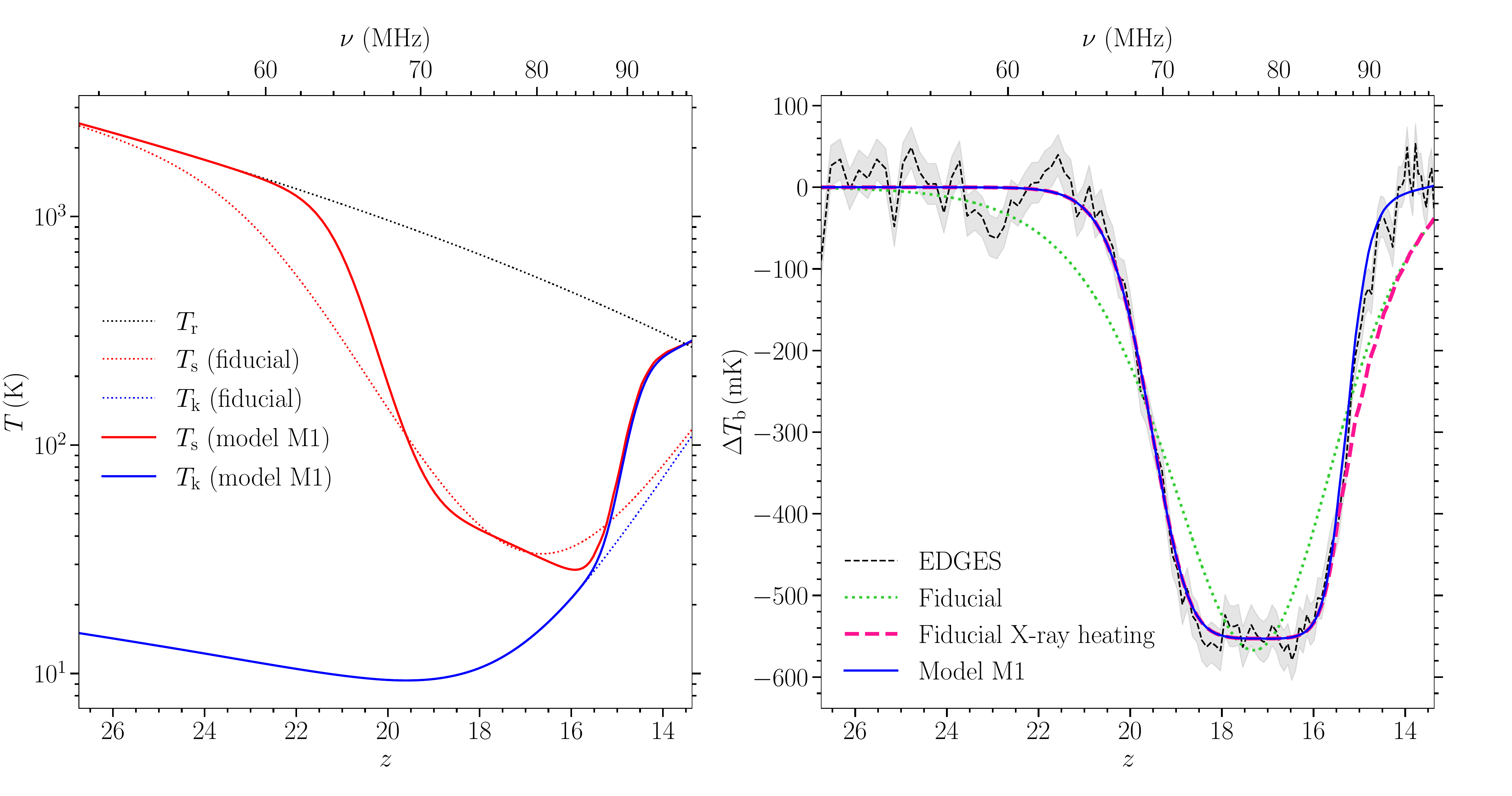}
\caption{Left panel: evolution of the radio background temperature (dotted black), the spin temperature (solid red) and the gas temperature (solid blue) in our model~M1. Right panel: the corresponding 21-cm signal (solid blue), in comparison with the EDGES measurement (dashed black), with a $\SI{50}{\milli\kelvin}$ uncertainty (grey band). The rms value of the residuals is $\SI{26}{\milli\kelvin}$, a dramatic improvement over the fiducial model. Dotted curves in both panels show the fiducial model of Figure~\ref{Fig1}. The dashed red curve in the right panel shows the case where the \lya\ emissivity follows model~M1 but the X-ray emissivity follows the fiducial model.}\label{Fig3}
\end{figure*}

Figure~\ref{Fig1} shows the cosmic thermal evolution (left panel) and the 21-cm signal (right panel) for the best-fitting values of the seven parameters of the model described above, when the model is fit to the EDGES measurement, assuming a constant $\SI{50}{\milli\kelvin}$ uncertainty across the frequency band \citep[see also][]{Hills, Mirocha19, Atrideb, hibbard2022}. The best-fitting parameter values are listed in Table~\ref{Tab1}. The Ly~$\alpha$ background is assumed to be set by a broken power law SED of Population II stars (metallicity $=10^{-3}$) assuming a \citet{scalo1998} initial mass function (IMF) \citep{Leitherer_1999, BL05}. The minimum virial temperature required is $\SI{e4.25}{\kelvin}$, equivalent to a circular velocity of $\SI{15.86}{\kilo\metre\per\second}$ and minimum halo mass of $1.65\times10^9\,\mathrm{M}_{\odot}h^{-1}$ (at $z=0$) assuming the mean molecular weight to be $\mu=1.22$, for which molecular hydrogen cooling fuels the star formation \citep{B16}. The X-ray background is assumed to be set by the high mass X-ray binaries; the normalization is about 2.4 times that suggested by extrapolating luminosity and SFR relation from \citet{Mineo} for 0.2--$\SI{30}{\kilo\electronvolt}$. The ERB required is $0.5$ per cent of CMB at $\nu_{\mathrm{obs}}=\SI{1420}{\mega\hertz}$, i.e., at $z=0$, which is very well below the upper limit of $9.6$ per cent of CMB -- $56.8$ per cent of ARCADE2/LWA1 \citep{Fixsen_2011, Dowell_2018} -- set by LOFAR \citep{Mondal}.

While the model is able to explain the central redshift and frequency of the observed absorption profile, the fit is quite poor at the falling and rising bounds of the absorption profile. The high-redshift `fall' in the profile lasts for about 65~Myr in our best-fitting model, while the low-redshift `rise' lasts for 97~Myr. (The corresponding numbers for the observed signal are 25~Myr and 51~Myr, respectively.) Indeed, overall the best-fitting model provides a very poor fit with a reduced-$\chi^2$ of $262/119$, which corresponds to a $p$-value (right tail probability) of just $8.7\times 10^{-13}$. [For our $\chi^2$ calculation we use spectrally flat errors $\SI{50}{\milli\kelvin}$ in magnitude motivated by noise level in the EDGES measurement \citep{Bowman}.] The rms value of the residuals relative to this model is $\SI{69}{\milli\kelvin}$.

In order to improve the fit at the falling edge of the profile, one seems to require a sudden drop in the spin temperature, either by a sudden evolution in the Ly~$\alpha$ coupling or by a drop in the gas temperature (for which no well-known mechanism exists). Similarly, the rapid rise in signal at $z\sim 15$ requires a sharp rise in the spin temperature, which is only possible by a corresponding rapid increase in the gas temperature [possibly by X-ray heating or exotic phenomenon such as cosmic ray heating \citep{Bera_22}].  

This problem persists even when we disregard the `non-standard' ERB. In the absence of an ERB, the model is no longer able to achieve the observed amplitude, but the shape and the central redshift would remain relatively similar to what is seen in Figure~\ref{Fig1}.

\begin{figure*}
\centering
\includegraphics[width=0.8\textwidth]{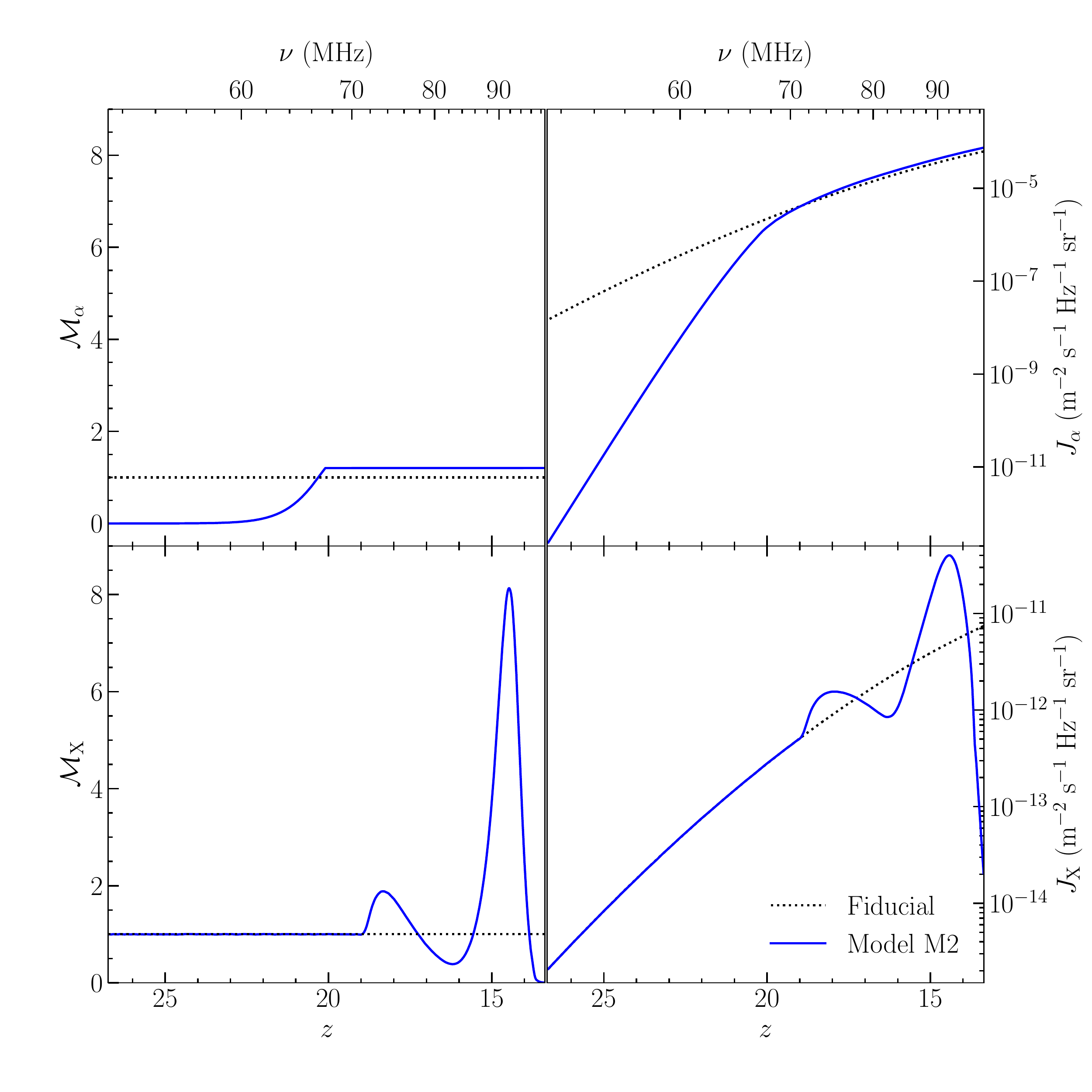}
\caption{Our model~M2. The left-hand side panels show the multiplicative factors $\ma$ and $\mx$ for the \lya\ and X-ray emissivity, respectively, required to fit the flat-bottomed shape of the EDGES measurement. The right-hand side panels show the corresponding background intensities $J_\alpha$ and $J_{\mathrm{X}}$. Dotted curves in all panels denote the fiducial model. This model presents the extreme scenario in which the \lya\ emissivity is held as close to the fiducial model as possible. In contrast to Figure~\ref{Fig2}, this model explains the flat bottom of the absorption profile by modifying the X-ray emissivity evolution.}\label{fx}
\end{figure*}

\begin{figure*}
\centering
\includegraphics[width=1\textwidth]{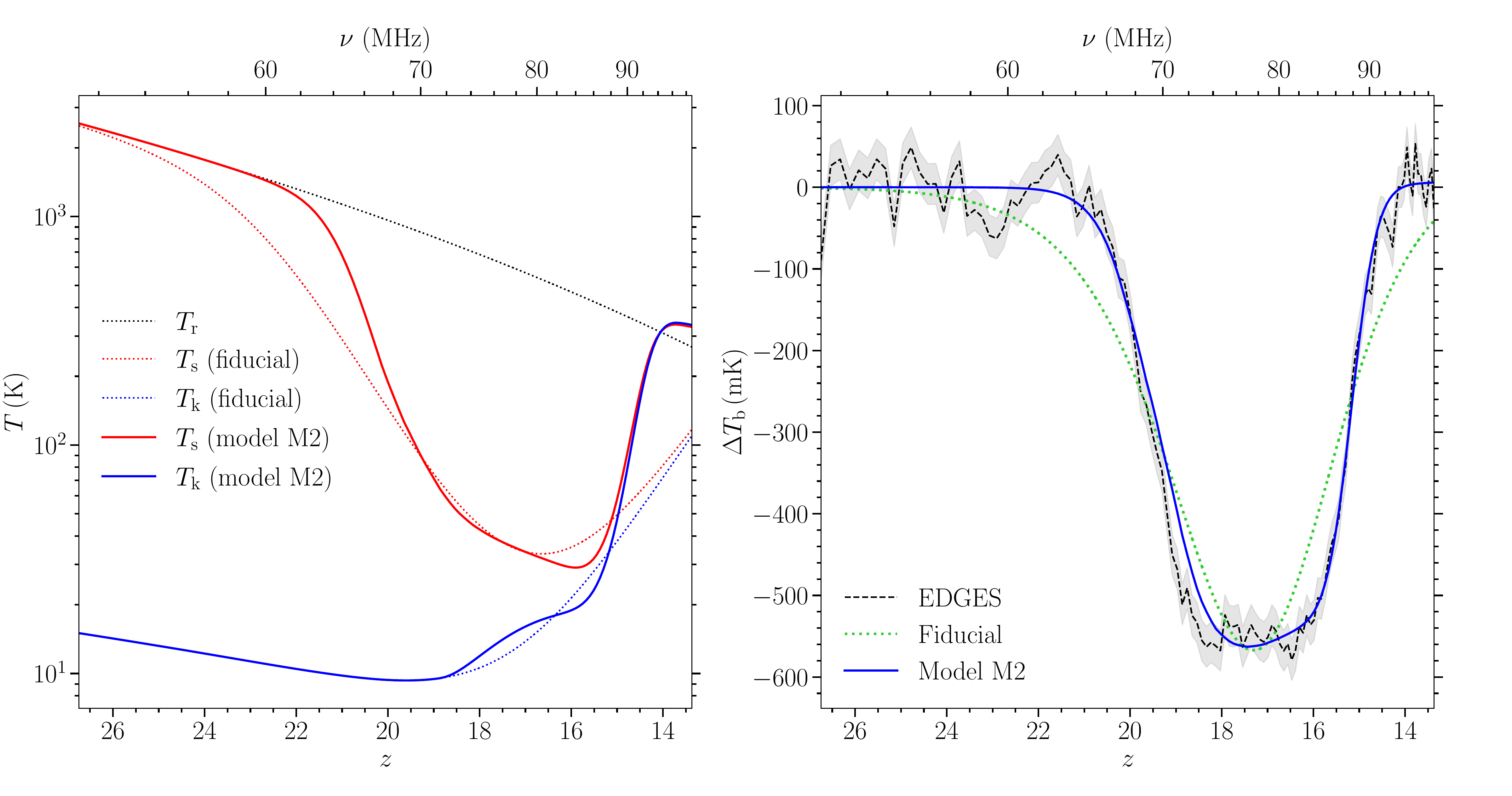}
\caption{Left panel: evolution of the CMB+ERB temperature (dotted black), the spin temperature (solid red) and the gas temperature (solid blue) in our model~M2. Right panel: the corresponding 21-cm signal (solid blue), in comparison with the EDGES measurement (dashed black), with a $\SI{50}{\milli\kelvin}$ uncertainty (grey band). The rms value of the residuals is $\SI{30}{\milli\kelvin}$, again a significant improvement over the fiducial model. Dotted curves in both panels show the fiducial model of Figure~\ref{Fig1}.}\label{mix}
\end{figure*}

\section{Direct consequences of a flat-bottomed 21-cm profile}\label{sec:new_models}

Thanks to the simplicity of Equations~\eqref{eqn:delta_tb} and \eqref{Ts}, the most direct implication of the flat-bottomed shape of the 21-cm absorption profile is a stringent requirement on the evolution of the \lya\ coupling and the gas temperature. A non-trivial evolution in one or both of these quantities is necessary if we are to improve the model fit shown in Figure~\ref{Fig1}. This non-standard evolution in $x_\alpha$ and $T_\mathrm{k}$ can in turn arise due to modified evolution of any of the several underlying processes, such as star formation, or \lya\ and X-ray emission. But before we investigate the implications for these more fundamental quantities, it is instructive to understand the primary implications for $x_\alpha$ and $T_\mathrm{k}$.

\subsection{\texorpdfstring{Ly~$\alpha$}{Ly~α} emission}\label{M1}

To understand the requirements for the Ly~$\alpha$ emission, starting from the best-fitting fiducial model described above, we can give a time dependence to Ly~$\alpha$ specific intensity, keeping everything else fixed, to try and reproduce the observed profile. Let the fiducial Ly~$\alpha$ background be $J_\alpha(z)$ and the required version be $J_\alpha'(z)$. We derive $J_\alpha'(z)$ as follows: the observed signal $\Delta T_\mathrm{b}^{\mathrm{obs}}$ corresponds to the spin temperature 
\begin{equation}
{T'_\mathrm{s}}^{-1}=T_\mathrm{r}^{-1}\left[1-\SI{5.29e-3}{\milli\kelvin^{-1}}\frac{\Delta T_\mathrm{b}^{\mathrm{obs}}}{h\Omega_\mathrm{b}(1-Y_{\mathrm{p}})\bar{x}_{\ion{H}{i}}}\sqrt{\frac{\Omega_\mathrm{m}}{1+z}}\right]\,,\label{spin_req}
\end{equation}
which is obtained by inverting Equation~\eqref{DeltaT}. Next, given Equation~\eqref{Ts}, we can obtain the required Ly~$\alpha$ coupling as
\begin{equation}
x_\alpha'=\frac{T_{\mathrm{r}}^{-1}-{T'_{\mathrm{s}}}^{-1}}{{T'_{\mathrm{s}}}^{-1}-T_{\mathrm{k}}^{-1}}\,.
\end{equation}
Since the Ly~$\alpha$ coupling is directly proportional to the \lya\ specific intensity we can write the required Ly~$\alpha$ specific intensity as
\begin{equation}
J'_{\alpha}=\frac{x_{\alpha}'}{x_{\alpha}}J_{\alpha}\,.\label{Jreq}
\end{equation}
Although theoretically it is easy to calculate $J'_{\alpha}$ by the above method, we must investigate if it is physically possible to have such a background. Stated more concretely, how should the Ly~$\alpha$ emissivity be modified to get $J'_{\alpha}$? We start by expressing the specific intensity in terms of emissivity as
\begin{equation}
\bar{J}_\alpha(z)=\frac{c}{4\pi}(1+z)^2\sum_{n=2}^{23}P_n\int_z^{z^{\mathrm{max}}_n}\mathcal{M}_\alpha(z')\frac{\epsilon_{\alpha}(E_n',z')}{H(z')}\,\ud z'\,,\label{Jobt}
\end{equation}
where $c$ is the speed of light and $\mathcal{M}_\alpha$ serves the purpose of providing $z$ dependence to the normalization parameter $f_\alpha$, so that $f_\alpha\to\mathcal{M}_\alpha(z) f_\alpha$. In the above equation $\epsilon_{\alpha}$ is the comoving emissivity defined as the number of photons emitted per unit comoving volume per unit proper time per unit energy and $P_n$ is the probability with which a photon in the upper Lyman lines will redshift to Ly~$\alpha$ wavelength.

Our task boils down to evaluating the modifying factor $\mathcal{M}_\alpha(z)$. Solving for $\mathcal{M}_\alpha$ is not straightforward because the integration limits in Equation~\eqref{Jobt} are functions of $z$. As a result, we can find this function by an iterative procedure in which we update $\mathcal{M}_\alpha(z)$ by a factor of $J_\alpha'/\bar{J}_\alpha$ after each iteration starting with $\mathcal{M}_\alpha(z)=1$.

We see that outside the absorption feature of EDGES, i.e. for $z>21$ and $15>z$, the signal is almost 0, which should imply the same for $\mathcal{M}_\alpha(z)$. At the centre of the profile, $z\sim17$, our current fitting already has the same depth as in the EDGES signal, which means we must have $\mathcal{M}_\alpha(z=17)\sim1$. At the high-redshift bound of the profile, from $z\sim18$ to 20, $\mathcal{M}_\alpha$ should be more than 1 and decrease to 0 towards higher redshifts. At the low-redshift bound of the profile, there is only a small amount of improvement that can be derived from the Ly$\alpha$ emissivity. This involves increasing $\mathcal{M}_\alpha$ until the Ly$\alpha$ coupling has saturated. Once the Ly$\alpha$ coupling has saturated, $T_\mathrm{s}=T_\mathrm{k}$, subsequent improvement can only be obtained by modifying the $T_\mathrm{k}$ evolution.

The modifying factor $\mathcal{M}_\alpha=\mathcal{M}_\alpha(z)$ obtained using the above iterative procedure is shown in Figure~\ref{Fig2} (top left panel). It has the shape we anticipated above. The black dotted line, $\mathcal{M}_\alpha=1$, represents the fiducial model. The top right panel in Figure~\ref{Fig2} shows the corresponding specific intensity (blue curve), in comparison with the fiducial model (black dotted curve). Incorporating the resultant specific intensity into our model leads to a significantly improved fit to the EDGES measurement, as shown in the right panel of Figure~\ref{Fig3} (dashed pink curve). The rms value of the residuals is $\SI{49}{\milli\kelvin}$, a dramatic improvement over the fiducial model.

In spite of this improvement, however, the model continues to perform poorly at low redshifts ($z\lesssim15.5$). The fit at $z\lesssim15.5$ in the dashed pink line can only be improved by increasing the gas temperature, which in our model can be done by enhancing the X-ray heating rate. For this purpose we introduce the function $\mathcal{M}_\mathrm{X}$, analogous to $\mathcal{M}_\alpha$, so that $f_\mathrm{X}\to \mathcal{M}_\mathrm{X}(z)f_\mathrm{X}$. Setting
\begin{equation}
\mathcal{M}_\mathrm{X}(z)=1+\ue^{-4(z-14.75)^2}\,,   
\end{equation}
improves the model to the solid blue line as seen in the right panel of Figure~\ref{Fig3}. The evolution of $\mathcal{M}_\mathrm{X}$ is shown in the bottom left panel of Figure~\ref{Fig2}, and the corresponding X-ray intensity is shown in the bottom right panel of this figure. The model corresponding to Figure~\ref{Fig2} has an rms value of the residuals $\SI{26}{\milli\kelvin}$, which is a dramatic improvement over the fiducial model. This model, which we label `M1', represents the direct consequence of the flat-bottomed shape of the 21-cm absorption profile for the \lya\ coupling, with minimal modification to the X-ray heating relative to the standard model. In the next section, we will examine if this requirement on the \lya\ coupling is plausible.  

It is worth stressing that in this model, the flat bottom of the absorption profile does not arise from a fine tuning of heating and cooling processes, as one might first guess. Instead, as we see in the left panel of Figure~\ref{Fig3}, the flatness arises out of a finely tuned non-standard evolution of the \lya\ coupling.  

\subsection{X-ray emission}\label{M2}

The above scenario with a maximum modification in Ly~$\alpha$ emissivity supplemented by a small change to X-ray emission for $z\lesssim14$ improves our model considerably. Here we consider an alternative scenario where we keep the modification to Ly~$\alpha$ background to a minimum, precisely for $z\gtrsim20$, but consider significant evolution of X-ray emission. The evolutionary factor $\mathcal{M}_\alpha(z)$ is now a monotonically increasing function for $z\gtrsim20$ (see upper left corner of Figure~\ref{fx}) which partly takes care of the sharp drop in the 21-cm signal. We design an X-ray background for $z\lesssim20$, which gives a flat bottom and a sharp rise in the 21-cm signal. We write the X-ray specific intensity in terms of X-ray emissivity modified by dimensionless function of $z$, $\mathcal{M}_\mathrm{X}$ as
\begin{equation}
\bar{J}_{\mathrm{X}}(E,z)=\frac{c(1+z)^2}{4\pi}\int_z^{z_{\star}}\mathcal{M}_\mathrm{X}(z')\frac{\epsilon_{\mathrm{X}}(E',z')}{H(z')}\ue^{-\tau_{\mathrm{X}}(E,z,z')}\ud z'\,,\label{Jx}
\end{equation}
where $z_\star\sim60$ is the redshift when the star formation starts \citep{Naoz}, and $\epsilon_{\mathrm{X}}$ and $\tau_{\mathrm{X}}$ are the X-ray emissivity and optical depth for X-ray photons, respectively. To find $\mathcal{M}_\mathrm{X}(z)$ we follow an iterative procedure as before: we update $\mathcal{M}_\mathrm{X}(z)$ by a factor of $J_{\mathrm{X}}'/\bar{J}_\mathrm{X}$ after each iteration starting with $\mathcal{M}_\mathrm{X}(z)=1$. Note that we should in principle choose $\mathcal{M}_\mathrm{X}$ to be a function of both energy and redshift. However, we find that it suffices to keep $E$ fixed at which $J_{\mathrm{X}}$ peaks, i.e., $E\sim \SI{0.2}{\kilo\electronvolt}$.

We first find the required form of X-ray specific intensity, $J_\mathrm{X}'$, which modifies the gas temperature evolution to result in a rapid increasing 21-cm brightness that is consistent with the EDGES measurement. We can estimate this from Equation~\eqref{tkeqn}. Since $q_{\mathrm{X}}\sim J_{\mathrm{X}}$, we get
\begin{equation}
J_\mathrm{X}'=\frac{(1+z)\ud T_\mathrm{k}'/\ud z-2T_{\mathrm{k}}'}{(1+z)\ud T_\mathrm{k}/\ud z-2T_{\mathrm{k}}}J_\mathrm{X}\,,
\end{equation}
where $T_\mathrm{k}'$ is the required gas temperature without altering the Ly~$\alpha$ coupling, which can be computed from
\begin{equation}
{T'_{\mathrm{k}}}^{-1}={T'_{\mathrm{s}}}^{-1}-\frac{T_{\mathrm{r}}^{-1}-{T'_{\mathrm{s}}}^{-1}}{x_\alpha}\,,
\end{equation}
where $T'_{\mathrm{s}}$ is the required spin temperature given in Equation~\eqref{spin_req}. The quantity $J_\mathrm{X}'$ so obtained may or may not be feasible. This assessment can be made by asking if there exists a function $\mathcal{M}_{\mathrm{X}}$ such that it gives $\bar{J}_\mathrm{X}=J_\mathrm{X}'$, where $\bar{J}_\mathrm{X}$ is defined in Equation~\eqref{Jx}.  Our iterative procedure answers this question.

Figure~\ref{fx} shows the resultant modification factors for the \lya\ and X-ray emissivities, alongside the corresponding specific intensities. Relative to Figure~\ref{Fig2}, $\ma$ now has small values, while $\mx$ has a relatively stronger evolution. The consequent thermal evolution and the 21-cm signal are shown in Figure~\ref{mix}. Similar to model M1, this model too is a significant improvement over the fiducial model, with a residual rms value of $\SI{30}{\milli\kelvin}$. This scenario represents the direct empirical implication of the EDGES measurement for the X-ray heating, with minimal modification to the fiducial \lya\ coupling. We term this model `M2'. Both models, M1 and M2, are much better at reproducing the observed shape of the 21-cm absorption profile. Unlike M1, however, model M2 does invoke a balance between heating and cooling to explain the flat bottom. This is achieved by means of a non-standard evolution of the X-ray heating rate. We will discuss the plausibility of this scenario in the next section.

It can be noticed that models M1 and M2 represent two extreme scenarios. M1 allows the \lya\ physics to deviate away from the fiducial picture but admits only a minimal modification to the X-ray model. Conversely, M2 allows the X-ray model to freely deviate away from the fiducial model but restricts modifications to the \lya\ physics to a minimum. One could possibly investigate the degeneracies between these two models, thereby constructing a family of models with flat-bottomed absorption profiles, but M1 and M2 already allow us to study the plausibility of an EDGES-like flat-bottomed absorption profile and outline its implications for high-redshift galaxies. 

\section{Astrophysical plausibility}\label{sec:plaus}

The factors $\mathcal{M}_\alpha$ or $\mathcal{M}_\mathrm{X}$ derived in previous sections are purely empirical. We now discuss if the values of these factors required by the EDGES measurement are reasonable. 

The non-trivial evolution of the \lya\ emissivity seen in Figure~\ref{Fig2} can be obtained either by a similarly non-trivial evolution in the cosmic SFRD or a change in the SED of the sources of UV photons. We will discuss modifications to the SFRD below. For now, it is worth asking if a standard SFRD evolution model could possibly result in the \lya\ emissivity required in our models M1 and M2. Previous literature worked with a fixed number of Lyman series photons per baryon at all epochs \citep{BL05}.
The cumulative Lyman series photons per baryon from newer population synthesis models such as \textsc{bpass} \citep[Binary Population and Stellar Synthesis;][]{eldridge, stanway} give somewhat higher values. For example, consider a stellar population formed in an instantaneous starburst, metallicity $Z=10^{-3}$ stars and a \citet{kroupa} IMF (a broken power law with slope $-1.3$ for masses $0.1$--$0.5\,\mathrm{M}_{\odot}$ and $-2.35$ for masses $0.5$--$300\,\mathrm{M}_{\odot}$).  This gives $\sim2\times10^4$ Lyman series photons per baryon implying $\ma\approx2.06$.

Recently, \citet{thomas} showed that Pop~III stars could have much brighter SEDs. A 2--$180\,\mathrm{M}_\odot$ power-law Pop~III IMF with slope $-0.5$ has $\ma\sim 6$. As a comparison, our extreme value for $\ma$ is $\sim 8$ at about $z=16$, as we see in Figure~\ref{Fig2}. It therefore appears that conventional Pop~II IMFs are unlikely to be able to produce the level of \lya\ emissivity demanded by the 21-cm signal unless, as we discuss below, the SFRD is boosted significantly. Pop~III IMFs, on the other hand, potentially can produce the required levels of \lya\ emission \citep[see also][]{Schauer_2019}. For instance, the burst feature in $\ma$ at $z=16$ could be the result of the peak in SED due to Pop~III stars of mass $\sim5\,\mathrm{M}_{\odot}$ \citep{thomas}.

Possible sources of X-rays include X-ray binaries (XRBs) \citep{Power_2013, Fragos_2013}, inverse Compton scattering in supernova remnants \citep{Oh_2001}, mini-quasars \citep{Madau_2004} and galactic winds \citep{Meiksin_2017}. Evidence suggests that quasars dominate at low redshifts while XRBs are dominant at high redshifts $(z\gtrsim5)$ \citep{Fabbiano, Brorby2016, Lehmer2016}. Among the two types of XRBs, high-mass XRBs are dominant over low-mass XRBs for $z\gtrsim2.5$ \citep{Fragos1}.

The X-ray SED $\phi_{\mathrm{X}}$ (defined in units of number per unit energy per stellar baryon, which for us would be such that $\phi_{\mathrm{X}}\propto E^{-5/2}$) is normalized by specifying $C_{\mathrm{X}}=\int E\phi_{\mathrm{X}}\,\ud E$ \citep{Mesinger11, Mesinger13}. The quantity $C_{\mathrm{X}}\equiv L_\mathrm{X}/\mathrm{SFR}$ is an observable that can be measured from the relationship between the X-ray luminosity $L_{\mathrm{X}}$ and the star formation rate (SFR). Measurements from nearby star-forming galaxies suggest a linear relationship, i.e., a constant $C_\mathrm{X}$ \citep{Grim03, Gilfanov}. In our previous work \citepalias{Mittal_2021} we adopted the following $L_{\mathrm{X}}$--SFR relation for photon energies 0.5--$\SI{8}{\kilo\electronvolt}$
\begin{equation}
C_{\mathrm{X,fid}}=\frac{L_{\mathrm{X}}}{\mathrm{SFR}}\approx\SI{2.61e32}{\watt}\left(\mathrm{M}_{\odot}\mathrm{yr}^{-1}\right)^{-1}\,,\label{LxSFR}
\end{equation}
based on observation of HMXBs in star-forming galaxies out to distances of $\lesssim\SI{40}{\mega\parsec}$ \citep{Mineo}. Observations of star-forming galaxies for $z\lesssim1.3$ \citep{Mineo2013} also suggest a scaling relation consistent with Equation~\eqref{LxSFR}. In \citetalias{Mittal_2021} and here, we extend the relation in Equation~\eqref{LxSFR} for photon energies 0.2--$\SI{30}{\kilo\electronvolt}$ \citep{Mirocha19}. The X-ray emissivity (in units of number per unit time per unit energy per unit comoving volume) is then given by
\begin{equation}
\epsilon_{\mathrm{X}}(E,z)=f_{\mathrm{X}}\phi_{\mathrm{X}}(E)\frac{\dot{\rho}_\star(z)}{m_{\mathrm{b}}}\,,\label{epsilonx}
\end{equation}
where $m_{\mathrm{b}}$ is the average baryon mass and $\dot{\rho}_\star$ is the comoving SFRD. The above emissivity then gives us the specific intensity as in Equation~\eqref{Jx}.

Our fiducial value of $f_{\mathrm{X}}$, as mentioned in Table~\ref{Tab1}, is $2.379$.  In the language of Section~\ref{M2}, this value of $f_\mathrm{X}$ corresponds to $\mx=1$, and Equation~\eqref{LxSFR} corresponds to $\mx=0.42$. At redshifts $z\gtrsim 1.3$, because of lack of direct observations, $f_{\mathrm{X}}$ remains largely unconstrained. However, evidence suggests that it is higher at higher redshifts \citep{Mirabel}. An evolving $f_{\mathrm{X}}$ (power law in $1+z$) was constrained by \citet{dij} based on the unresolved cosmic X-ray background \citep{Lehmer_2012} assuming that this background is produced by HMXBs. But this was applicable to much lower redshifts $(z\lesssim2)$ compared to the redshifts of our interests. At higher redshifts $(z\gtrsim5)$ the unresolved cosmic X-ray background puts only weak constraints on $f_{\mathrm{X}}$ \citep[see also][]{Fialkov_16}.

Not much is known about $f_\mathrm{X}$ for $z>14$. The only information for these redshifts we have is by \citet{Fragos_2013} which is based on population synthesis models. They gave fitting functions for HMXB luminosity density for $0\leqslant z\leqslant20$. Converting this to $C_{\mathrm{X}}$ we find that $C_{\mathrm{X}}$ peaks at $z\approx1.7$, decreases to about 0.025 at $z\approx19$ before increasing slightly to $\sim 0.03$ for $z\sim 20$.  Translating this to our language, $\mx(z)$ varies from about 0.7 at $z<2$ to $0.013$ at $z\sim 20$, far smaller than what is required by the EDGES measurement (in either of our models M1 and M2, Figures~\ref{Fig2} and \ref{fx}).  It thus appears that obtaining rapidly increasing X-ray background of the sort required by data is not plausible in conservative models.  One way of increasing $\mx$ is to again invoke Pop~III HMXBs, which can increase $\mx$ to values as large as $10^3$ \citep{Mirocha18}.

It is worth pointing out that the low-redshift spikes in $\ma$ and $\mx$ in Figures~\ref{Fig2} and \ref{fx} are connected to the rapid disappearance of 21-cm signal around $z\sim 15$.  The onset of reionization can also potentially explain this rapid disappearance.  Although reionization at such high redshifts will violate CMB constraints, such a scenario could potentially lead to a picture in which $\ma$ and $\mx$ level off or continue rising at $z\lesssim 15$.

\begin{figure*}
\centering
\includegraphics[width=1\textwidth]{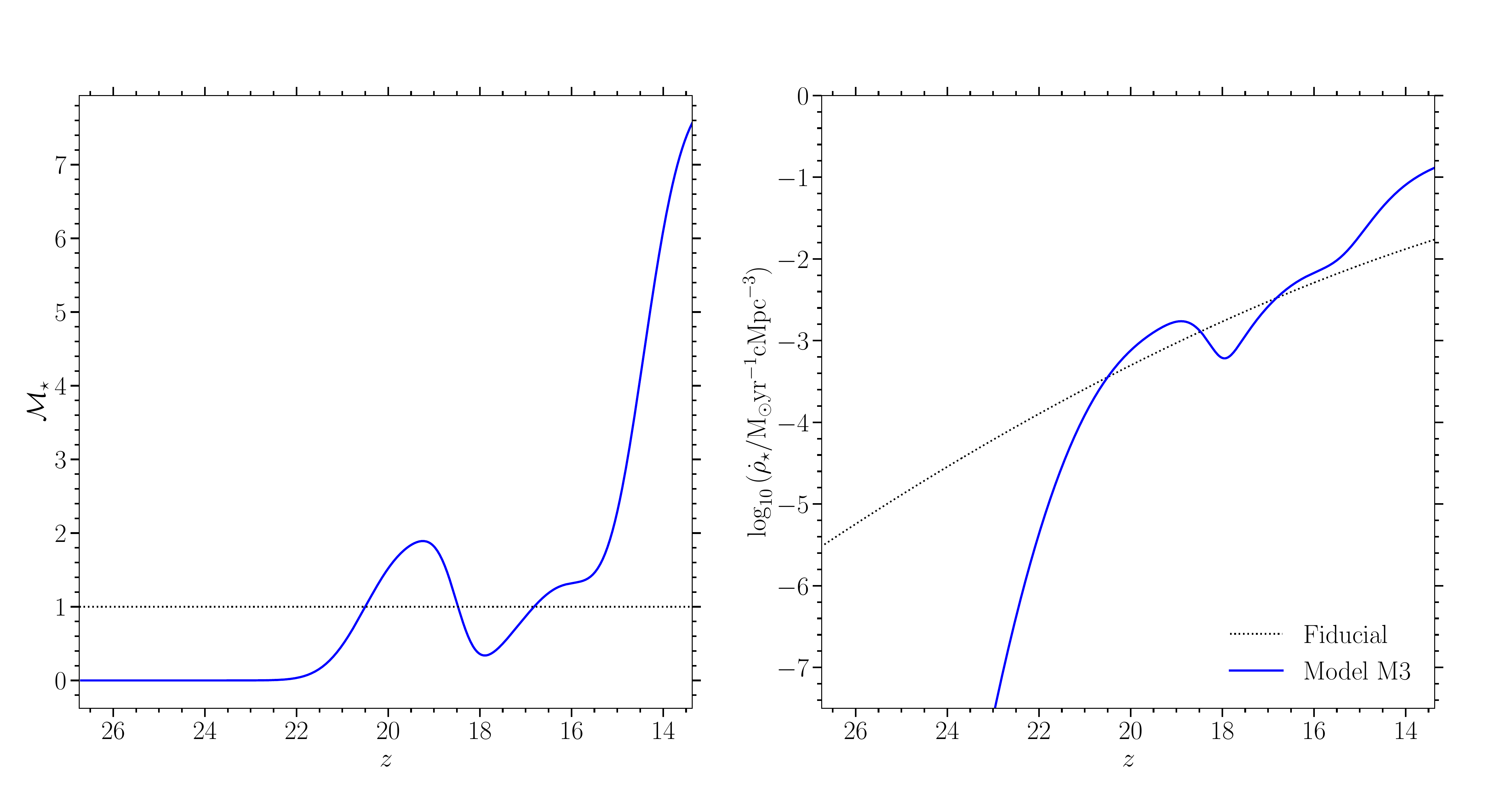}
\caption{Our model~M3. The left panel shows the $z$-dependent multiplication factor $\mstar$ required to fit the flat-bottomed shape of the EDGES measurements with our fiducial value for stellar properties.  The right panel shows the corresponding comoving cosmic star formation rate density (SFRD).  The black dotted line in both panels shows our fiducial SFRD, which has $f_{\star}=0.1$.}\label{sfrd}
\end{figure*}

\begin{figure*}
\centering
\includegraphics[width=1\textwidth]{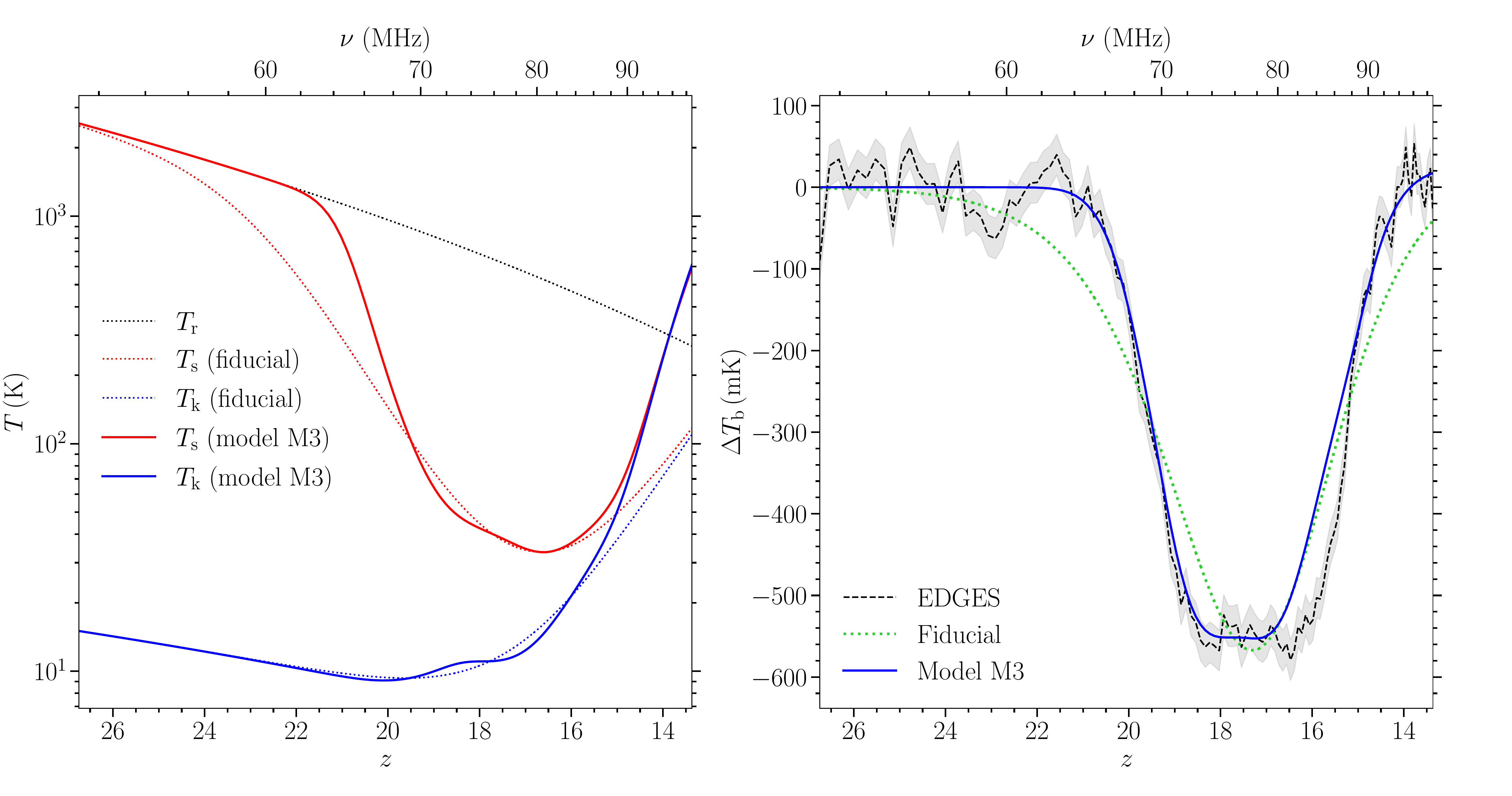}
\caption{Left panel: evolution of the CMB+ERB temperature (dotted black), the spin temperature (solid red) and the gas temperature (solid blue) in our model~M3. Right panel: the corresponding 21-cm signal (solid blue), in comparison with the EDGES measurement (dashed black), with a $\SI{50}{\milli\kelvin}$ uncertainty (grey band). The rms value of the residuals is $\SI{47}{\milli\kelvin}$.  The green dotted curve is the fiducial model of Figure~\ref{Fig1}.}\label{star}
\end{figure*}

\section{Implications}\label{sec:implic}

\subsection{Implications for high-redshift star formation}

We see above that the EDGES measurement demands \lya\ and X-ray production mechanisms that are at the limits of what is reasonable.  An alternative therefore is that the cosmic star formation evolution is modified relative to our fiducial model.  

In \citetalias{Mittal_2021} we worked with a constant value of SFE (defined as the fraction of gas in the dark matter haloes which gets converted into stars) of $f_\star=0.1$. However, in general $f_\star$ depends on feedback mechanisms, metallicity of the collapsing gas, and the mass of the star forming haloes \citep{Cohen17, Monsalve_2019}. In order to obtain a flat-bottomed absorption profile, we follow a similar method as before and change $f_\star$ so that it now has a time dependence: $f_\star\to \mathcal{M}_\star(z) f_\star$. (We continue to hold the SFE independent of the halo mass.  Also, the values of other model parameters, including $f_\alpha$ and $f_\mathrm{X}$ are held fixed to the fiducial values listed in Table~\ref{Tab1}.) Changing $f_\star$ inevitably affects both \lya\ as well as X-ray backgrounds.  As a result, the calculation of $\mathcal{M}_\star$ is not as straightforward as that of $\ma$ and $\mx$ in Models M1 and M2. We use hit-and-trial method for this starting with $\mathcal{M}_\star=\mathcal{M}_{\alpha}$.  The result is shown in Figure~\ref{sfrd}.

We see that the 21-cm profile requires the SFE to drop rapidly to zero at redshift higher than $\sim 20$.  This is to prevent a \lya\ background from getting established too early.  A rapid enhancement followed by a quick drop in $f_\star$ relative to the fiducial model is then required at redshift of $z\sim 20$--$19$ to reproduce the sharp falling edge of the profile. Finally, the model requires a significant enhancement in $f_\star$ once more at $z\lesssim 17$ to obtain the required X-ray heating. The resulting signal is a good fit, as shown in Figure~\ref{star}. This model has a residual rms value of $\SI{47}{\milli\kelvin}$.

This is a considerable improvement over the fiducial model.  But this model is still significantly worse than the models M1 and M2 in which the properties of the \lya\ and X-ray sources were varied. Thus it is hard to fit the EDGES data for $17>z>15$ by only modifying the SFRD. This can be understood to be because X-ray and Ly~$\alpha$ have contrasting effects on the 21-cm signal. Increasing SFRD for $17>z>15$ to get more Ly~$\alpha$ and hence a better fitting (by lower the signal) also increases X-ray heating at the same time (in turn raising the signal). The effects are reversed if we consider lowering the SFRD. Thus, if we leave the intrinsic properties of Ly~$\alpha$ and X-ray sources untouched and modify only the SFRD as shown in Figure~\ref{sfrd} then the signal shown in Figure~\ref{star} is the best that can be achieved.

It is also possible to get an equivalent model by changing the minimum mass of star-forming haloes. This minimum mass in our fiducial model corresponds to a virial temperature of $\SI{e4.25}{\kelvin}$, which is close to the limit set by atomic hydrogen cooling. The required reduction in the SFRD at $z\gtrsim 18$ necessitates a very high value of the minimum halo mass. Conversely, the large SFRD required at the lower redshift end of the data can be obtained by reducing the minimum mass by a factor of $\lesssim 2$.

\subsection{Predictions for galaxy surveys}\label{galaxy}

Given the required SFRD evolution, and with our assumption that the SFE is independent of the halo mass, we can make predictions for the ultraviolet luminosity functions (UV LFs) of high-redshift galaxies. As a measure of the comoving number density of galaxies for a given luminosity, the UV LF has a one-to-one correspondence with the halo mass function. The LF is traditionally expressed as the comoving number density per unit absolute magnitude rather than number density per unit luminosity. We construct the intrinsic LF as
\begin{equation}
\mathrm{LF}=\frac{\ud \varphi(M_{\lambda})}{\ud M_{\lambda}}=\frac{\ud n(M_{\mathrm{h}})}{\ud M_{\mathrm{h}}}\frac{\ud M_{\mathrm{h}}}{\ud L_\lambda}\frac{\ud L_\lambda}{\ud M_\lambda}\,,\label{main}
\end{equation}
where $\ud n$ is comoving number density of haloes with mass between $M_{\mathrm{h}}$ and $M_{\mathrm{h}}+\ud M_{\mathrm{h}}$, $L_\lambda$ is the specific luminosity at a wavelength $\lambda$, and $M_\lambda$ is the corresponding absolute magnitude. We consider the UV LF for $\lambda=\SI{1500}{\angstrom}$.

\begin{figure*}
\centering
\includegraphics[width=1\textwidth]{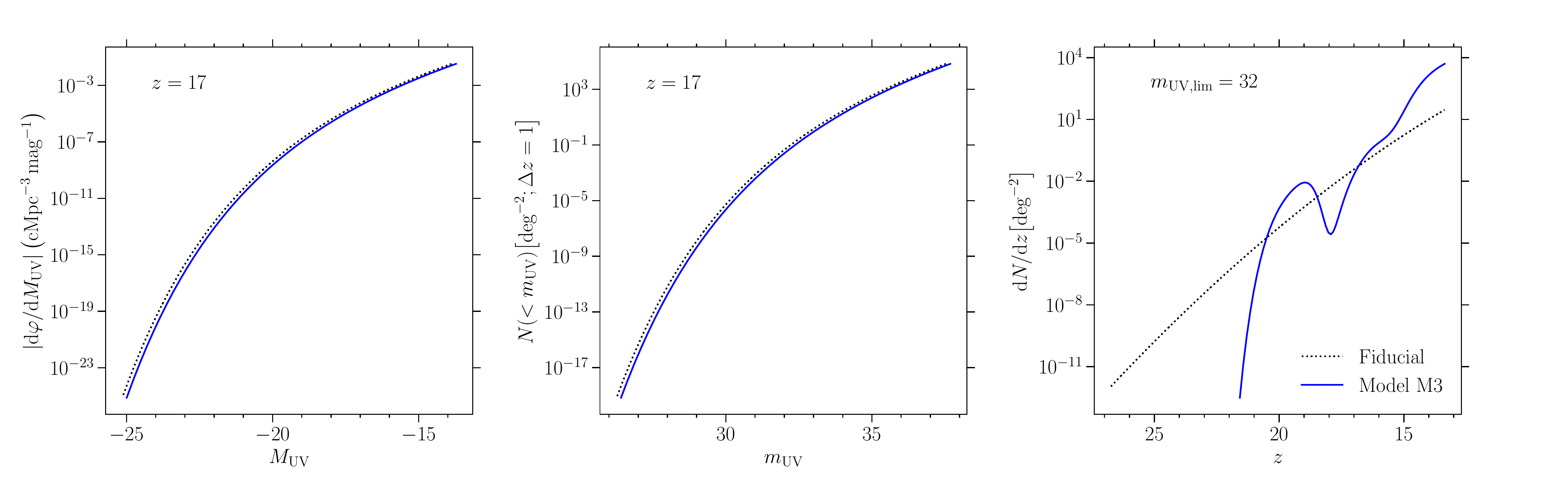}
\caption{Left panel: galaxy UV luminosity function at $z\sim 17$ for $M_\mathrm{UV}<-13$. Middle panel: number density of $z=17$ sources on the sky. Right panel: predictions for an ultra-deep \textit{JWST} survey with $m_{\mathrm{UV,lim}}=32$. In all panels, dotted curves show our fiducial models; solid blue curves show model~M3.}\label{phiuv}
\end{figure*}

The UV luminosity of a given star-forming halo can be assumed to track the instantaneous SFR \citep[e.g.,][]{ken, dick} so that 
\begin{equation}
L_\mathrm{UV}=\dot{M}_\star\mathcal{L}_{\mathrm{UV}}\,,\label{L}
\end{equation}
where $\dot{M}_\star$ is the SFR and $\mathcal{L}_{\mathrm{UV}}=\SI{8.695e20}{\watt\hertz^{-1}}\big(\mathrm{M}_\odot \mathrm{yr}^{-1}\big)^{-1}$ is the specific luminosity per unit SFR \citep{Mirocha17}. We follow \citet{sun16} to model the SFR as
\begin{equation}
\dot{M}_\star=f_\star \dot{M}_{\mathrm{b,acc}}\,,\label{sfr}
\end{equation}
where $\dot{M}_{\mathrm{b,acc}}$ is accretion of baryons onto dark matter haloes. A halo's baryonic mass accretion rate can be described by the empirical relation \citep{mcbride, sun16, Mirocha17}
\begin{equation}
\dot{M}_{\mathrm{b,acc}}=\dot{M}_{0}\left(\frac{M_{\mathrm{h}}}{10^{10}\mathrm{M}_{\odot}}\right)^a\left(\frac{1+z}{7}\right)^b\,,\label{bmar}
\end{equation}
where $\dot{M}_{0}=3\,\mathrm{M}_{\odot}$yr$^{-1}$, $a=1.127$ and $b=2.5$. Combining equations~\eqref{L}, \eqref{sfr} and \eqref{bmar} we get
\begin{equation}
L_\mathrm{UV}=f_\star\dot{M}_{0}\left(\frac{M_{\mathrm{h}}}{10^{10}\mathrm{M}_{\odot}}\right)^a\left(\frac{1+z}{7}\right)^b\mathcal{L}_{\mathrm{UV}}\,.\label{L1}
\end{equation}
The corresponding apparent magnitude (in the AB system) is given by \citep{Oke74}
\begin{equation}
m_\mathrm{UV}=-2.5\log_{10}\left[\frac{1}{4\pi}\left(\frac{L_\mathrm{UV}}{\SI{1}{\watt\hertz^{-1}}}\right)\left(\frac{ d_{\mathrm{L}}}{\SI{1}{\metre}}\right)^{-2}\right]-56.1\,,\label{mAB}
\end{equation}
where
\begin{equation}
d_{\mathrm{L}}(z)=c(1+z)\int_0^z\frac{\ud z'}{H(z')}\,,\label{ld}
\end{equation}
is the luminosity distance. The corresponding absolute magnitude is
\begin{equation}
M_\mathrm{UV}=-2.5\log_{10}\left[\frac{1}{4\pi}\left(\frac{L_\mathrm{UV}}{\SI{1}{\watt\hertz^{-1}}}\right)\left(\frac{ d_{10}}{\SI{1}{\metre}}\right)^{-2}\right]-56.1\,,\label{MAB}
\end{equation}
where $d_{10}=\SI{10}{\parsec}$.

If we assume that $f_\star$ is independent of $M_{\mathrm{h}}$, then using equations~\eqref{L1} and \eqref{MAB} we arrive at
\begin{equation}
\frac{\ud\ln M_{\mathrm{h}}}{\ud M_\mathrm{UV}}=-\frac{2\ln10}{5a}\,.
\end{equation}
Inserting the above into Equation~\eqref{main} we get
\begin{equation}
\mathrm{LF}=-\frac{2\ln10}{5a}\frac{\ud n}{\ud \ln M_\mathrm{h}}\,,\label{lf}
\end{equation}
where we will use the \citet{Press} form for the mass function $\ud n/\ud\ln M_\mathrm{h}$. Because the above relation gives us the LF as a function of halo mass we again use equations~\eqref{L1} and \eqref{MAB} to eliminate $M_{\mathrm{h}}$ to get LF as a function of $M_{\mathrm{UV}}$.  As a consistency check, we used the above formalism to verify that our model results in a UV LF consistent with observations from \citet{Bouwens_2015}.


For high-redshift galaxy surveys, such as those planned with the \textit{JWST} or \textit{Nancy Grace Roman Space Telescope} (\textit{NGRST}), a useful quantity that can be derived from the LF is the galaxy count brighter than a certain limiting luminosity, $L_{\mathrm{UV,lim}}$, per unit redshift per unit angular area on the sky. This can be defined as \citep{white_2010}
\begin{equation}
\frac{\ud N}{\ud z}=\frac{\ud V}{\ud z}\int_{L_{\mathrm{UV,lim}}}^{\infty}\frac{\ud\varphi}{\ud L_{\mathrm{UV}}}\ud L_{\mathrm{UV}}\,,\label{count1}
\end{equation}
where the differential comoving volume element per square radian is given by comoving area on the sky per square radian times the comoving differential length, i.e., \citep{hogg}
\begin{equation}
\ud V=\left[(1+z)^2\,d^2_\mathrm{A}\right]\left[(1+z)\,c\,\ud t\right]=\frac{c}{H(z)}\left(\frac{d_\mathrm{L}}{1+z}\right)^2\ud z\,,
\end{equation}
where $d_\mathrm{A}$ is the angular diameter distance. Equation~\eqref{count1} simplifies to
\begin{equation}
\frac{\ud N}{\ud z}(M_{\mathrm{h,lim}},z)=\frac{c}{H(z)}\left(\frac{d_\mathrm{L}}{1+z}\right)^2\int_{M_{\mathrm{h,lim}}}^{\infty}\frac{\ud n}{\ud M_{\mathrm{h}}}\ud M_{\mathrm{h}}\,,
\end{equation}
where $M_{\mathrm{h,lim}}$ is the limiting halo mass which gives luminosity $L_{\mathrm{UV,lim}}$. Finally, to get galaxy count brighter than a limiting luminosity per unit angular area on the sky we evaluate $N(M_{\mathrm{h,lim}})=\int(\ud N/\ud z) \ud z$.

Figure~\ref{phiuv} shows our predictions. The left panel shows our prediction for the UV LF at $z=17$ for two values of the SFE. Since in our modelling SFE depends only on $z$ and not on halo mass, different curves are shifted horizontally by just a constant. Using our best-fitting model parameters we show the UV LF by a black dotted line for which SFE is 0.1, while solid blue curve corresponds to a modified SFE, about 0.09 (as $\mathcal{M}_{\star}(z=17)\approx0.9$). The middle panel of Figure~\ref{phiuv} shows the sky surface density of $z=17$ galaxies. Note that the absolute (apparent) AB magnitude range in the figure, $-25<M_{\mathrm{UV}}<-14\, (26<m_{\mathrm{UV}}<38)$, is mapped to halo mass between $M_{\mathrm{h}}=10^{12}\,\mathrm{M}_{\odot}$ and $10^8\,\mathrm{M}_{\odot}$. The absolute and apparent magnitude corresponding to minimum star forming halo virial temperature are $-12.5$ and $38.9$ at $z=17$, respectively.

The right panel of Figure~\ref{phiuv} shows the evolution of the sky surface density given a fixed limiting flux for which the apparent magnitude is $m_\mathrm{UV,lim}=32$, corresponding to an ultra-deep survey using \textit{JWST} \citep{Mason_2015}. The dotted curve in this figure corresponds to our fiducial SFRD evolution, which requires extreme modification to the \lya\ and X-ray SEDs (via the $\ma$ and $\mx$ factors of our models M1 and M2). The solid curve shows the case corresponding to Figure~\ref{sfrd}, in which the stellar population is held fixed to have rather conservative choices for the SED but an EDGES-like flat-bottomed absorption profile is obtained via a non-trivial modulation of the SFRD evolution. This SFRD evolution is imprinted in the observable $\mathrm{d}N/\mathrm{d}z$.

It is interesting to note that the models shown in Figure~\ref{phiuv} do not modify the shape of the UV LF, but only its evolution. This can be compared with the prediction by \citet{Mirocha19}, who studied models that also modified the shape of the LF \citep[see also][]{Madau_18}. Nonetheless, it is interesting that \citet{Mirocha19} also find that the EDGES measurement requires an order-of-magnitude increase in galaxy number counts relative to canonical extrapolations.

\begin{figure}
\centering
\includegraphics[width=1\linewidth]{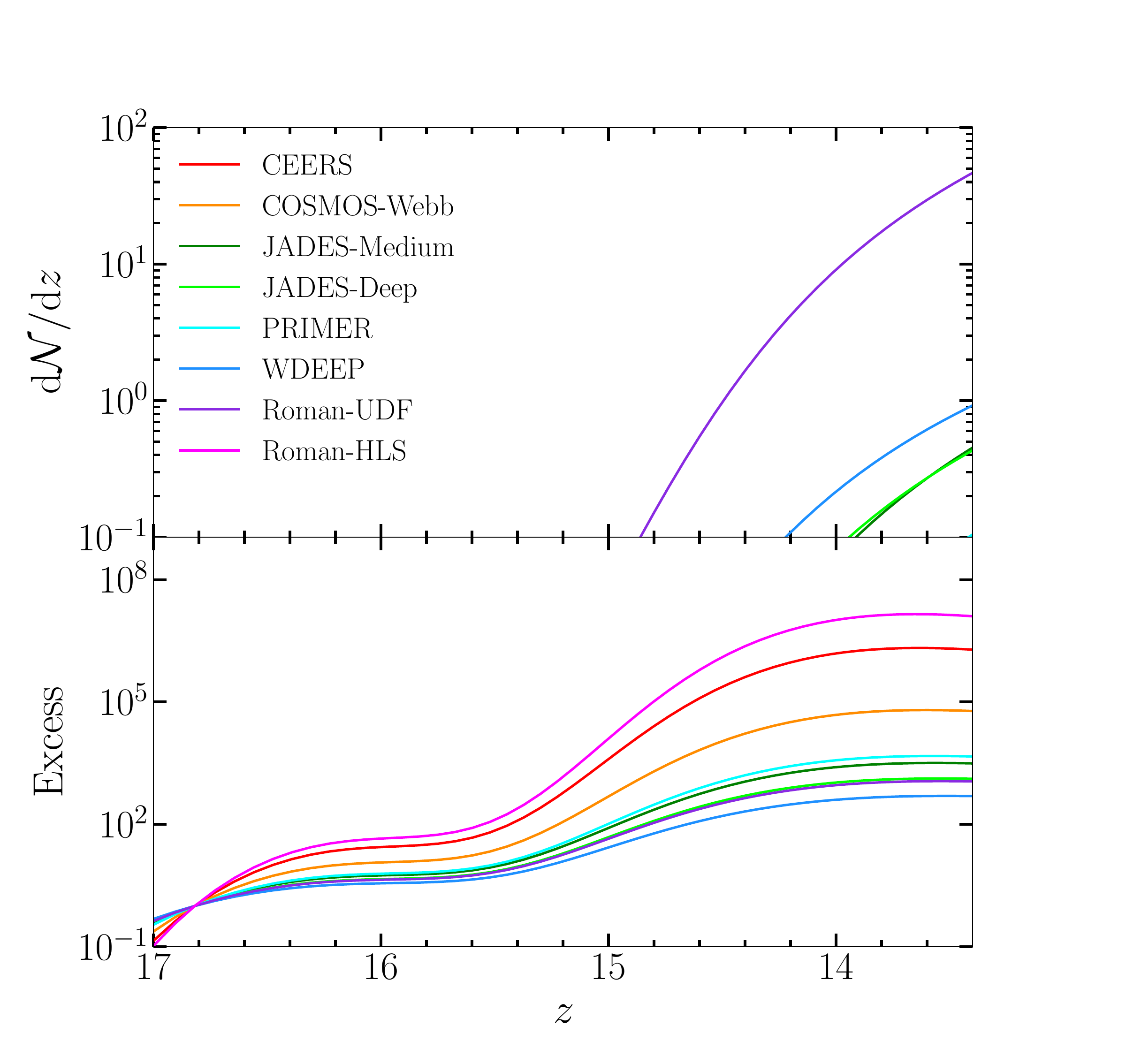}
\caption{Top panel: $\mathcal{N}'$ for model M3. Bottom panel: $\mathcal{N}'/\mathcal{N}'_{\mathrm{fiducial}}$, where $\mathcal{N}'=\ud \mathcal{N}/\ud z$ is redshift gradient of the cumulative count of galaxies brighter than the survey limit. The survey areas and limiting magnitudes are listed in Table~\ref{Tab2}.}\label{fig:surveys}
\end{figure}

Figure~\ref{fig:surveys} presents our results for surveys soon to be underway on the \textit{JWST} and two surveys proposed using the upcoming \textit{NGRST}. The top panel in Figure~\ref{fig:surveys} shows our predicted $\ud \mathcal{N}/\ud z$ for model M3 as a function of redshift, where $\ud \mathcal{N}/\ud z$ is the redshift gradient of the cumulative count of galaxies brighter than the survey limit. The bottom panel shows the excess relative to our fiducial model, i.e., $$\left(\frac{\ud \mathcal{N}}{\ud z}\right)\cdot\left(\frac{\ud \mathcal{N}}{\ud z}\right)^{-1}_{\mathrm{fiducial}}\,.$$ We have shown results for CEERS \citep{ceers}, COSMOS-Webb \citep{cosmos}, JADES-Medium, JADES-Deep \citep{robertson}, PRIMER \citep{primer}, WDEEP \citep{WDEEP}, Roman-UDF \citep{rudf}, and Roman-HLS \citep{hls}. The survey areas and limiting magnitudes are listed in Table~\ref{Tab2}.  WDEEP, JADES-Deep, and JADES-Medium on the \textit{JWST}, and Roman-UDF on the \textit{NGRST}, have the most optimistic configurations for detecting the enhancements in galaxy counts predicted here.


\begin{table}
\centering
\caption{Surveys with \textit{JWST} (first six) and the Nancy Grace Roman Space Telescope (last two) considered in our analysis, with their area and limiting UV magnitude (AB system).}\label{Tab2}
\begin{tabular}{llc}
\hline
Name  & Area & Limit. mag. $(m_{\mathrm{UV}})$\\\hline
\rule{0pt}{2.5ex}CEERS & $100\,{\mathrm{arcmin}^2}$ & 26.5\\
COSMOS-Webb & $0.6\,{\mathrm{deg}^2}$ & 27.7\\
JADES-Medium & $290\,{\mathrm{arcmin}^2}$ & 29.3\\
JADES-Deep & $46\,{\mathrm{arcmin}^2}$ & 29.9\\
PRIMER & $144\,{\mathrm{arcmin}^2}$ & 29.0\\
WDEEP & $10.8\,{\mathrm{arcmin}^2}$ & 30.8\\
Roman-UDF & $1\,{\mathrm{deg}^2}$ & 30.0\\
Roman-HLS & $2000\,{\mathrm{deg}^2}$ & 26.0 \\ \hline
\end{tabular}
\end{table}

\section{Conclusion}\label{sec:conc}

The cosmological 21-cm absorption signal measured by the EDGES collaboration has a distinctly different shape than predictions from standard models. In this paper, we have developed models that provide far superior fits to the EDGES measurement than standard models. These new models either invoke exotic stellar populations or modulate the cosmic SFRD evolution in order to establish a cosmic Ly~$\alpha$ background within 25\,Myr from $z\sim 21$ to 19, and a cosmic X-ray background within 50\,Myr from $z\sim 16$ to 14.

Scenarios involving exotic stellar populations have the advantage of the cosmic SFRD evolution in these models being relatively standard.  Our models M1 and M2 presented in Section~\ref{sec:new_models} are two examples of such models. However, as discussed in Section~\ref{sec:plaus} above, these models seem to require extreme Population~III IMFs evolving  over redshifts. It appears difficult to obtain the required level of both \lya\ as well as X-ray emission from models with standard stellar populations.

On the other hand, in scenarios involving modifications to the cosmic SFRD, more conservative assumptions about stellar populations are sufficient. The model~M3 presented in Section~\ref{sec:implic} is an example. This models the cosmic star formation to rapidly drop to zero at redshifts higher than $z \sim 21$. This leads to robust predictions for upcoming deep galaxy surveys using \textit{JWST} or \textit{NGRST}. Combining data from such surveys with global 21-cm measurements should be a rich source of information about galaxy formation as well as radiative background at cosmic dawn.

Our three models represent extremes of their respective categories. Model~M1 admits large modifications to the \lya\ source properties but attempts to keep the properties of X-ray sources to their conventional values. Conversely, model~M2 tries to get the maximum contribution from the X-ray side. Model~M3 retains standard assumptions for the \lya\ and X-ray sources and maximises the contribution from SFRD modifications.

There are obvious degeneracies between these models. These degeneracies could perhaps be captured by developing parametric phenomenological models of the radiative sources and the cosmic SFRD evolution. Indeed, without such extensions to the standard models, it would not be possible to fit the EDGES measurement to a satisfactory level at all. This would also prevent inferring constraints on exotic physics models affecting the thermal properties of the universe at cosmic dawn. One way to develop such extensions is to consider families of fitting functions to the \lya\ and X-ray emissivities discussed in Section~\ref{sec:new_models}. Another possibility is to fold such possibilities in end-to-end forward models of the signal such as those employed by the REACH experiment \citep{Eloy}.

\section*{Acknowledgements}

We thank Basudeb Dasgupta, Nicole Drakos, Anastasia Fialkov, Thomas Gessey-Jones, Avery Meiksin, Jordan Mirocha, Brant Robertson, Saurabh Singh, and Sandro Tacchella for helpful discussions, and Joop Schaye for editorial comments that improved this paper. It is also a pleasure to acknowledge discussions with members of the REACH (Radio Experiment for the Analysis of Cosmic Hydrogen) collaboration. GK is partly supported by the Department of Atomic Energy (Government of India) research project with Project Identification Number RTI~4002 and by the Max Planck Society through a Max Planck Partner Group.

\section*{Data availability}

The data and code underlying this article will be shared on reasonable request to the corresponding author.

\bibliographystyle{mnras}
\bibliography{Biblo}

\appendix

\bsp
\label{lastpage}
\end{document}